\documentclass[mathleft
]{an}
\usepackage{graphicx}
\usepackage{times}
\begin{document}

\Pagespan{1}{}
\Yearpublication{ }%
\Yearsubmission{2010}%
\Month{ }%
\Volume{ }%
\Issue{ }%

\title{The black hole in NGC 1313 X-2}

\author{Luca Zampieri\inst{1}\fnmsep\thanks{Corresponding author: \email{luca.zampieri@oapd.inaf.it}\newline}
\and Alessandro Patruno\inst{2}
}
\titlerunning{The Black Hole in NGC 1313 X-2}
\authorrunning{L. Zampieri \& A. Patruno}
\institute{
INAF-Osservatorio Astronomico di Padova, Padova I-35122, Italy
\and 
Astronomical Institute 'A. Pannekoek', University of Amsterdam, Science Park 904, 1098 XH Amsterdam, The Netherlands
}

\received{ }
\accepted{ }
\publonline{ }


\abstract{%
The amount of data available for NGC 1313 X-2
make it a cornerstone for the study of ultraluminous X-ray sources (ULXs).
We modelled the optical and X-ray data of this ULX with a binary evolution code that takes into account
X-ray irradiation. We restricted the candidate binary system to be
either a 50--100$M_\odot$ black hole (BH) accreting from a 12--15$M_\odot$ main sequence star
or a 20$M_\odot$ BH with a 12--15$M_\odot$ giant donor. If the orbital period
of the system is $\sim$6 days, a 20$M_\odot$ BH model becomes unlikely and we
are left with the only possibility that the compact accretor in NGC1313 X-2
is a massive BH of 50--100$M_\odot$. We briefly discuss these results within
the framework of an alternative scenario for the formation of ULXs, in which
a portion of them may contain BHs of $\ga 30-40 M_\odot$ formed from very
massive stars in a low metallicity environment.
}

\maketitle

\section{Introduction}

Several pieces of observational evidence strongly suggest that a large fraction of 
ULXs are accreting black hole X-ray binaries with massive 
donors (see, e.g., Zampieri \& Roberts \cite{zr09} and references therein). Their X-ray spectral properties,
along with the long-term flux variability, are similar to those observed in Galatic X-ray
binaries although, in the last few years, {\it Chandra} and {\it XMM-Newton} observations
revealed new behaviours, showing the existence of spectral states with rather peculiar 
properties (e.g. Gladstone et al. \cite{getal09}). At the same time, while most ULXs appear to show little variations
on timescales of seconds to hours (e.g. Heil et al. \cite{hetal09}), in some cases fast X-ray 
variability is reminiscent of the behaviour observed in Galactic accretion-powered sources
(rms variability, broad band noise, quasi periodic oscillations;
e.g. Strohmayer \& Mushotzky \cite{sm03}, Strohmayer et al. \cite{setal07}, Feng et al. \cite{fetal10}, Heil \& Vaughan \cite{hv10}). Probably, the most compelling 
evidence of the binary nature of ULXs is provided so far by the detection of a modulation in the X-ray 
light curve of M 82 X-1 ($\sim$ 62 days; Kaaret et al. \cite{ketal06}) and NGC 5408 X-1 ($\sim 115$ days;
Strohmayer \cite{s10}), that are interpreted as the orbital periods of these systems.

A number of ULXs are associated to stellar optical counterparts and some of them have been
identified with stars of known spectral type (e.g. Liu et al. \cite{liu02,liu04}, Kaaret et al. \cite{kaaret04}, 
Mucciarelli et al. \cite{mucciarelli05}, Soria et al. \cite{soriaetal05}).
The counterparts appear almost ubiquitously hosted in young stellar environments 
(e.g. Pakull et al. \cite{pakull06}, Ramsey et al. \cite{ramsey06}, Liu et al. \cite{liu07}) and have properties consistent with those of young, 
massive stars. NGC 1313 X-2 is one of the few ULXs with a well identified optical counterpart. 
It is located $\sim$~6' south to the nucleus in the barred spiral galaxy NGC 1313 
at a distance of  3.7--4.27 Mpc (Tully \cite{tully88}, M\'endez et al. \cite{mendezetal02}, Rizzi et al. \cite{rizzietal07}). 
Its observed X-ray luminosity varies between a few $\times 10^{39}$ erg s$^{-1}$
and $\sim 10^{40}$ erg s$^{-1}$ in the 0.3--10 keV band (Feng \& Kaaret \cite{feng06}, \cite{mucciarelli07}).
The large amount of data available make this source a cornerstone for the study 
of ULXs.

\section{A bit of history: X-rays}

NGC 1313 X-2 was discovered by {\it Einstein} (Fabbiano \& Trinchieri \cite{ft87})
and included in the {\it Einstein Extended Medium Sensitivity Survey} as MS 0317.7-6647.
Stocke et al. (1995) originally proposed that it could be either a Galactic isolated neutron star or a
binary containing a massive BH in NGC 1313.

It was then repeteadly observed with other satellites, including {\it ROSAT} 
(1991--1998; Stocke et al. \cite{s95}, Colbert et al. \cite{c95}, Miller et al. \cite{m98}, Schlegel et al. \cite{s00}), {\it ASCA} (1993--1995; 
Petre et al. \cite{p94}, Makishima et al. \cite{m00}), 
{\it Chandra} and {\it XMM-Newton} (2000--2006; Zampieri et al. \cite{z04}, Feng \& Kaaret \cite{fk05} and references
therein). The X-ray spectrum of NGC 1313 X-2 can be modelled 
in different ways and shows the typical 
features shared by other bright ULXs, such as the presence of a soft, thermal component
and a rollover of the spectrum above $\sim 3$ keV (e.g. Gladstone et al. \cite{getal09}, Stobbart et al. \cite{s06} 
and references therein).
The amount of {\it XMM-Newton} spectra now available allow for a tentative identification
of different spectral states (see Pintore \& Zampieri, these Proceedings).

\section{A bit of history: optical}

The optical counterpart of NGX 1313 X-2 was first identified on an ESO 3.6 m $R$ band image thanks
to an accurate {\it Chandra} astrometry and after registering the X-ray image 
on the position of SN 1978K (Zampieri et al. \cite{z04}). ESO VLT images of the field showed
that the counterpart was actually composed of two distinct objects (C1 and C2),
separated by $\sim 0.7$" (Mucciarelli et al. \cite{mucciarelli05}). Further refinement in the
astrometry and accurate modelling of the optical emission indicated that object C1 was the 
more likely counterpart (Liu et al. \cite{liu07}, Mucciarelli et al. \cite{mucciarelli07},
Patruno \& Zampieri \cite{pz08}). However, this was established 
beyond any doubt by the detection of the 4686 He II emission line in its optical
spectrum, a characteristic imprint of X-ray irradiation (Pakull et al. \cite{pakull06}, 
Gris\'{e} et al. \cite{grise08}).

The stellar environment of NGC 1313 X-2 has also provided interesting constraints. 
There are two groups of young stars spread out over $\sim 200$ pc.
Isochrone fitting of the colour-magnitude diagram of these groups has been attempted and 
provides cluster ages of 20$\pm$5 Myrs (Pakull et al. \cite{pakull06}, Ramsey et al. \cite{ramsey06}, 
Liu et al. \cite{liu07}, Gris\'{e} et al. \cite{grise08}).
As several other ULXs, NGC 1313 X-2 is also associated with a very extended 
($\sim$ 400 pc) optical emission nebula that gives important information on the 
energetics and lifetime of the system (Pakull \& Mirioni \cite{pakull02}). Assuming that it is
formed in an explosive event or that its mechanical energy comes from the ULX 
wind/jet activity, the characteristic energy and age of the nebula turn out to be 
$\sim 10^{52}-10^{53}$ erg and $\sim$1 Myr, respectively (Pakull et al. \cite{pakull06}).
Finally, several different estimates of the mass and luminosity class of the optical
counterpart of NGC 1313 X-2 were reported in the literature (see e.g. Patruno \& Zampieri \cite{pz10} and 
references therein). 

Recently, Liu et al. (\cite{liu09}) found a possible periodicity of 6.12$\pm$0.16 days in the B band
light curve of the optical counterpart of NGC 1313 X-2, that was interpreted as the orbital period
of the binary system. Three cycles were detected in the $B$ band, while no modulation was found in $V$.
Previous studies carried out on the available HST and VLT observations led to negative results
(Gris\'{e} et al. \cite{grise08}).
More recently, lack  of significant photometric variability on a new sequence of VLT
observations has been reported by Gris\' et al. (2009).
In principle, the detection of the orbital period would definitely confirm
the identification of the optical counterpart and the binary nature of this system.
Most importantly, it would open the way to perform a dynamical measurement of 
the BH mass.

\section{Optical constraints}

In the following we use the X-ray luminosity and all the available optical data on NGC 1313 X-2 to constrain the 
properties of this ULX and its BH. This is probably one of the most accurate analyses
that is possible to perform at present on a single ULX. We adopt the $V$ and $B$ band
photometry of the counterpart as determined by Mucciarelli et al.
\cite{mucciarelli07}. As the source is variable, we have further corrected
the magnitudes and colours by using the average value
of $V$ and by propagating the errors on $V$ and $B$. The
error in the absolute magnitudes is taken to be equal
to the maximum uncertainty in the different distance
determinations of NGC 1313 (Tully \cite{tully88}, M\'endez et al. \cite{mendezetal02}, Rizzi et al. \cite{rizzietal07}).

Concerning the reddening, two different estimates of the colour excess 
were adopted in the literature: $E(B - V ) = 0.1$ (\cite{mucciarelli07}, 
Gris\'{e} et al. \cite{grise08})
and $E(B - V )=0.3$ (Liu et al. \cite{liu07}). In the following we chose
$E(B-V )=0.11$. If we assume $E(B-V)=0.3$, it is not possible
to obtain agreement between the binary evolution 
models and all the simultaneous constraints coming from observations.
For $E(B-V ) = 0.11$ the adopted absolute $V$ magnitude is therefore in
the range -4.38 to -4.79 and the $B-V$ colour is $-0.13 \pm 0.06$.

Three different values of the metallicity inferred from an abundance analysis 
of the HII regions in NGC 1313, all sub-solar, are reported in the literature: 
$Z=0.1 Z_\odot$ (Pilyugin \cite{p01}), $Z=0.2 Z_\odot$ (Ryder \cite{r93}) and $Z=0.4 Z_\odot$ (Walsh \& Roy \cite{wr97},
Hadfield \& Crowther \cite{hc07}).
Therefore, although the metallicity in the environment of NGC 1313 X-2 appears to be 
definitely sub-solar, the actual value is still uncertain (see Ripamonti et al., 
these Proceedings; see also Pintore \& Zampieri, these Proceedings, for an
independent metallicity estimate from X-ray spectra).

Finally, we re-analyzed the optical photometry of NGC 1313 X-2 with the aim of 
clarifying the statistical significance of the orbital periodicity identified by 
Liu et al. (\cite{liu09}). As reported in Impiombato et al. (these Proceedings), 
the folded $B$ band light curve shows a 6 days periodicity with 
a significance slightly larger than $3\sigma$. This suggests that the periodicity 
may be there, but the low statistical significance of the $B$ band modulation, along 
with the lack of detection in the $V$ band, make its identification uncertain.

\begin{figure}
\includegraphics[width=80mm]{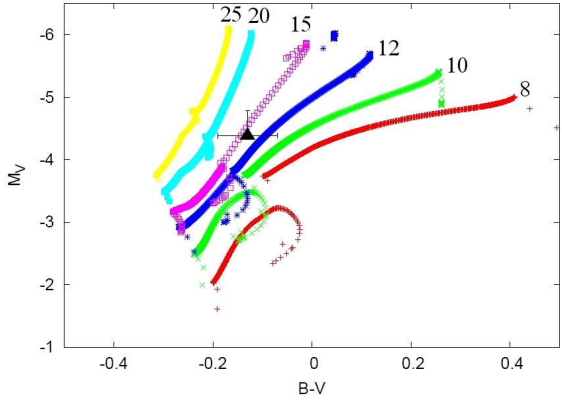}
\caption{CM diagram for binaries with a 20$M_\odot$ BH undergoing {\it case
AB} mass transfer. The black triangle marks the position of the optical counterpart
of NGC 1313 X-2. All the tracks are plotted only during the 
contact phases. The lower part of each track (up to the turn around) refers to the MS RLOF phase,
whereas the upper part to the H-shell burning RLOF phase.
The labels on each curve indicate that mass of the donor star (in $M_\odot$).}
\label{fig1}
\end{figure}

\begin{figure}
\includegraphics[width=80mm]{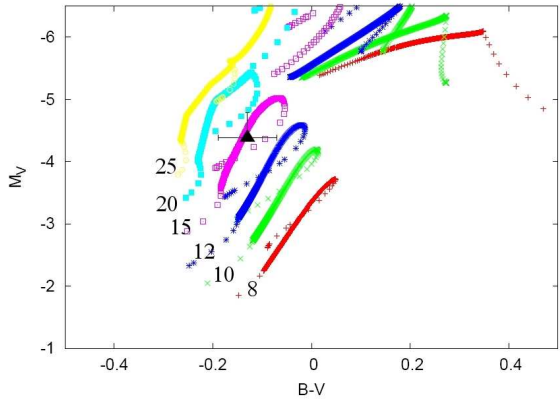}
\caption{Same as Figure~\ref{fig1} for a 70$M_\odot$ BH ({\it case AB}).}
\label{fig2}
\end{figure}

\section{Modeling the NGC 1313 X-2 binary system}

We modelled NGC 1313 X-2 using the binary evolution code 
described in Patruno \& Zampieri (\cite{pz08}), that computes the evolution of the orbital parameters
and the companion star taking into account the contribution of the accretion 
disc and X-ray irradiation during the contact phases. For this analysis, we developed
a new version of the code, with a more accurate calculation of the optical colours (Patruno \& Zampieri \cite{pz10}). 

The evolution is computed using two different values of the (zero age) metallicity, 
both sub-solar: $Z = 0.2 Z_\odot$ and $Z = 0.5 Z_\odot$. We found that the effects on the evolutionary tracks and
the stellar colours are minimal. Hence, for reference, in the following we will consider 
only tracks computed for $Z=0.5 Z_\odot$.

If the donor starts a contact phase via
Roche-lobe overflow (RLOF), we assume that a standard accretion disc forms around the BH. 
The accretion rate is instantaneously taken to be equal to the
mass-transfer rate from the companion. When the accretion rate $\dot M$
exceeds the Eddington rate $\dot M_{Edd}$, we impose $\dot M = \dot M_{Edd}$ and assume
that the excess mass is expelled from the system. In our simulations,
all the binaries that start RLOF on the main sequence (MS) have
also a second episode of mass transfer after the terminal age main
sequence (TAMS). Following Patruno \& Zampieri (\cite{pz10}), we term such binary models as {\it Case
AB}, while those starting the first contact phase after the TAMS are
termed {\it Case B}. For the latter, typically the contact phase is too short 
when compared to the characteristic age of the nebula ($\sim$~1 Myrs), which
provides an estimate of the active phase of the ULX. Therefore, from now on, we will
focus on {\it Case AB} systems.


\subsection{Evolution of a binary with a $20M_\odot$ BH}

Figure~\ref{fig1} shows the evolutionary tracks on the color-magnitude (CM) diagram 
computed for binary systems with a BH of $20 M_\odot$ and 
donors of different masses. Considering the intrinsic variability of the counterpart,
its position on the CM diagram may be broadly in agreement with $10-20 M_\odot$ donors 
during H-shell burning. However, only stars in the $12-15 M_\odot$ have characteristic 
ages consistent with the estimated parent cluster age (15-25 Myrs). Donors this massive,
during H-shell burning, provide a mass transfer rate largely above Eddington. 
In this case there is excess mass that must be somewhat expelled from the system. 
Beaming and/or super-Eddington emission is necessary to explain the observed ULX luminosity.
We note also that the system certainly underwent a first contact phase during 
MS, during which it had time to energize the surrounding nebula.

\subsection{Evolution of a binary with a $70M_\odot$ BH}

Figure~\ref{fig2} shows the evolutionary tracks on the CM diagram for a systems with 
a BH of $70 M_\odot$ and donors of different masses. The position
of the optical counterpart is now consistent with the tracks of stars on the MS, 
in the mass range between 12 and 20$M_\odot$. Again, the further constrain inferred
from the parent cluster age (15-25 Myrs) restricts the allowed donor masses to the interval
$12-15M_\odot$. The mass transfer of a 15$M_\odot$ at this stage is typically between
$2\times 10^{-7}$ and $2\times 10^{-6}$ $M_\odot$ yr$^{-1}$, consistent with the 
accretion rate ($\sim 10^{-6}$ $M_\odot$ yr$^{-1}$) inferred from the measured 
average luminosity of NGC 1313 X-2 ($\sim 4 \times 10^{39}$ erg s$^{-1}$; Mucciarelli et al. \cite{mucciarelli07}).
Similar results are obtained for BHs with a mass between 50 and 100 $M_\odot$.

\subsection{Evolution of the orbital period}

The evolution of the orbital period $P$ is rather different for the two systems.
In order to have a first contact phase during MS, the initial orbital separation
is $\sim 1$ day. For the 20$M_\odot$ BH binary, the orbital period at the end of 
MS is $\sim 4$ days. During the H-shell burning phase it increases rapidly up to 
20 days.
When the $12-15 M_\odot$ donor tracks cross the 
position of the counterpart on the CM diagram, $P\ga 8$ days.

For the 70$M_\odot$ BH binary, the orbital period at TAMS reaches $\sim 6$ days
and is between 5.5 and 6.5 days when the evolutionary tracks of the $12-15 M_\odot$ 
donors are consistent with the photometric properties of the optical counterpart.

\section{Discussion}

According to the actual value of the orbital period of the system and considering
the constraints from the inferred mass transfer rate, the position on the CM diagram, 
the characteristic ages of the parent stellar cluster and the bubble nebula, the 
following scenarios are possible:
\begin{itemize}
\item $P\la 4$ days: mass transfer from a $\sim 15 M_\odot$ MS donor onto a $\sim 20 M_\odot$ BH,
if all the mass is accreted onto the BH and super-Eddington emission is allowed.
\item $P=5-7$ days: mass transfer of a MS donor of $12-15 M_\odot$ onto a $50-100 M_\odot$ BH.
\item $P\ga 8-10$ days: mass transfer of a H-shell burning donor of $12-15 M_\odot$ onto 
a $\sim 20 M_\odot$ BH. If $P\sim 12$ days, the photometry is consistent with
the possition of the counterpart only for donors of $\sim 20 M_\odot$, too young 
with respect to the age of the parent cluster.
\item $P\ga 15-20$ days: mass transfer of a H-shell burning donor of $\sim 15 M_\odot$ onto 
a $\sim 20 M_\odot$ BH, only if X-ray irradiation is switched off (because, e.g., emission 
is beamed and not able to hit the disc and donor surfaces).
\end{itemize}

If we use the tentative identification of the $\sim 6$ days orbital period originally
reported by Liu et al. (\cite{liu09}) as a further constraint, the observational data appear
to be consistent only with the second scenario and we are left with the only possibility 
that the compact accretor in NGC 1313 X-2 is a massive black hole of $50-100 M_\odot$.

Such a system would fit well within a massive BH scenario for ULXs forming in a 
low-metallicity natal environment (Mapelli et al. \cite{mapelli09}, Zampieri \& Roberts
\cite{zr09}, Mapelli et al. \cite{mapelli10}; Mapelli et al., 
these Proceedings). Albeit uncertain, in fact, the metallicity of the environment of
NGC 1313 X-2 is definitely sub-solar. As mass loss becomes less efficient in such
environments, already at $Z\la 0.2-0.3 Z_\odot$ massive stars may retain a significant fraction of their envelopes 
at the end of their life (see e.g. Zampieri \& Roberts \cite{zr09} and references therein).
If the envelope is more massive than $\sim 30-40 M_\odot$, a low metallicity star may 
collapse directly to form a BH (Heger et al. \cite{h03}, Belczynski et al. \cite{b09}). This occurs because,
after core collapse, the supernova shock wave loses too much energy in trying to unbind 
the envelope until it stalls and most of the star collapses into a BH of mass comparable 
to the star final mass ($\ga 30-40 M_\odot$; Fryer \cite{f99}, Zampieri \cite{z02}). With these massive BHs
only very modest beaming or slight violations of the Eddington limit (a factor 
of a few) are needed to account for the luminosity of bright ($\ga 10^{40}$ erg s$^{-1}$) ULXs.

As noted above, however, the identification of the orbital period of NGC 1313 X-2 remains
uncertain. Furthermore, we did not perform a complete survey of the parameter space evolving 
{\it case AB} systems with BH masses between 20 and 50$M_\odot$.
For BH masses in this range there may also be agreement with observations with
other values of $P$. Finally, modelling the amplitude and shape 
of the light curve (considering X-ray irradiation and ellipsoidal modulations)
may give additional constraints. A systematic investigation of this type is 
postponed to when a more robust assessment of the orbital period will be available.

\acknowledgements
LZ acknowledges financial support through \\ INAF grant PRIN-2007-26. We thank
the referee for useful comments. AP acknowledges  support from
the Netherlands Organization for Scientific Research (NWO) Veni Fellowship.


\end{document}